\RequirePackage{fix-cm}
\documentclass[twocolumn,epjc3]{svjour3}  
\smartqed  
\RequirePackage{graphicx}
\usepackage{subfigure}
\usepackage{booktabs}
\usepackage{multirow}
\usepackage{amssymb}

\journalname{Eur. Phys. J. C}

\begin{document}

\title{On the Possibility of Measuring the Single-tagged Exclusive Jets at the LHC}


\author{Maciej Trzebi\'nski\thanksref{e1}
        \and
        Rafa{\l} Staszewski
        \and
        Janusz Chwastowski
}

\thankstext{t1}{This work was supported in part by Polish National Science Centre under contract 2012/05/B/ST2/02480.}
\thankstext{e1}{Corresponding author: maciej.trzebinski@ifj.edu.pl}


\institute{H. Niewodnicza\'nski Institute of Nuclear Physics Polish Academy of Sciences\\
ul. Radzikowskiego 152, 31-342 Krak\'ow, Poland. \label{addr1}
}

\date{Received: date / Accepted: date}

\newcommand{\dd}{\mathrm{d}}

\maketitle

\begin{abstract}
The feasibility studies of the measurement of the central exclusive jet production at the LHC using the proton tagging technique are presented. In order to reach the low jet-mass region, single tagged events were considered. The studies were performed at the c.m. energy of 14 TeV and the ATLAS detector, but are also applicable for the CMS-TOTEM experiments. Four data-taking scenarios were considered: AFP and ALFA detectors as forward proton taggers and $\beta^* = 0.55$~m and $\beta^* = 90$~m optics. After the event selection, the signal-to-background ratio ranges between 5 and $10^{4}$. Finally, the expected precision of the central exclusive dijet cross-section measurement for data collection period of 100 h is estimated.
\end{abstract}

\section{Introduction}
Diffraction has always been an important part of the studies performed in experiments involving hadron interactions. This is true also for the LHC, where a large community works on both theoretical and experimental aspects of possible diffractive measurements. 

\begin{figure}
  \centering
  \includegraphics[width=0.45\columnwidth]{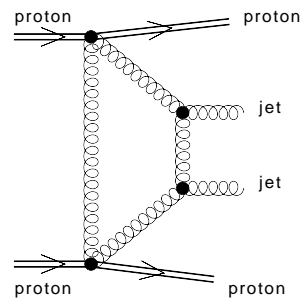}
\caption{Central exclusive jet production: both interacting protons stay intact and two jets are produced.}
\label{fig_exc_jj_diag}
\end{figure}

In the majority of collisions at the LHC interacting protons break-up and their remnants are populating forward regions. However, in a fraction of events the protons interact coherently, either electromagnetically -- by exchanging a photon, or strongly -- via an exchange of a colour singlet object named Pomeron. In such situation, called a diffractive production, one or both protons stay intact, lose part of their energy and are scattered at very small angles into the accelerator beam pipe. The central exclusive production (CEP) consists a special class among diffractive processes. In these events both protons stay intact and all energy available due to the colourless exchange is used to produce the central system. A diagram of central exclusive jet production is shown in Figure~\ref{fig_exc_jj_diag}. Such events were observed at the Tevatron~\cite{exc_jj_cdf_D0} and are also expected to happen at the LHC. Their measurement will be an important test for the applicability of QCD for such processes and can serve as a discrimination tool for phenomenological models.

There are few theoretical descriptions of the exclusive jet production available. In this paper the model developed by Khoze-Martin-Ryskin (KMR)~\cite{KMR} is used. In the KMR model a perturbative approach is used -- the colourless exchange is represented by an exchange of two gluons; a hard and a soft one. The role of the soft gluon is to provide the colour screening that ensures that no net colour charge is exchanged between the two protons. The exclusivity is assured via a Sudakov
form factor~\cite{Sudakov}, which suppresses the radiation of additional gluons. The two-gluon exchange and lack of additional radiation means that the central exclusive dijet production is extremely rare in comparison to standard dijet production at the LHC. In addition, the dijet system is produced in a $J^{PC} = 0^{++}$ state.

The exclusive jets can be also produced as a result of photon-photon interaction. However, as this process is of electromagnetic nature, the expected cross-section is much smaller than the one for estimated within the KMR approach. Thus it is not further considered.

In order to perform a fully exclusive measurement there is a need to measure both the jets and the intact protons. The requirement of both protons being tagged often forces a production of large amount of energy in the central region which, in turn, significantly reduces the cross-section. In consequence, there is a need to collect a large amount of data ($\mathcal{O}($fb$^{-1})$) and to operate in high pile-up environment. Such measurement is feasible (see Ref.~\cite{ATLAS_exclusive}) but very challenging. In this paper we discuss the possibility of the measurement of the central exclusive jet production at the LHC at the centre-of-mass energy of $\sqrt{s} = 14$ TeV in cases when only one intact proton is measured.

\section{Monte Carlo Generators}
The KMR model is embedded in the \textsc{FPMC} generator~\cite{FPMC}. \textsc{FPMC} is a modification of \textsc{Herwig 6.5}~\cite{Herwig} and uses the final state parton showering and hadronisation algorithms included in that generator. For exclusive processes the initial state parton showers are forbidden and accounted for as a part of the calculation. Due to the exclusivity requirements all energy lost by interacting protons is transfered into the produced central system.

The diffractive backgrounds: double Pomeron exchange (DPE) and single diffractive (SD) jets were also generated using \textsc{FPMC}, assuming the rapidity gap survival factor of 0.03 and 0.1, respectively. The generation of these events is based on the the resolved Pomeron model \cite{Ingelman} and makes use of H1 2007 Fit B~\cite{Hera_fit}. The multi-parton interactions (MPI) were turned off.
The non-diffractive (ND) jets were generated using \textsc{Pythia8}~\cite{Pythia8}.

Minimum-bias events were generated using \textsc{Pythia8} with MBR tune~\cite{MBR}. The following processes were taken into account: non-diffractive production, elastic scattering, single diffractive dissociation, double diffractive dissociation and central diffraction.

Jets were reconstructed using anti-$k_T$ algorithm with the jet radius $R = 0.6$ as implemented in the \textsc{FastJet} package~\cite{FastJet}. As an input for the analysis, all generated final state particles were considered.

Generated protons were transported to the location of the considered forward detector using the \textsc{FPTrack} program~\cite{FPTrack} -- a tool dedicated for fast proton tracking through the LHC magnetic structures. After the transport the proton energy was reconstructed using a procedure described in Ref.~\cite{unfolding}.

\section{Experimental Environment}
Exclusive jet production events could be selected by looking for rapidity gaps in the forward direction or by measuring the forward protons. This paper focuses on the proton tagging technique. The studies were performed for the ATLAS detector~\cite{ATLAS} case (with two sets of forward proton detectors: ALFA and AFP), but the conclusions are also valid for the similar set of detectors installed around the CMS/TOTEM Interaction Point~\cite{TOTEM}.

The ALFA (Absolute Luminosity For ATLAS) detector system consists of four detector stations placed symmetrically with respect to the ATLAS IP at 237 m and 245 m~\cite{ALFA}. In each ALFA station there are two Roman Pot devices allowing vertical movement of the detectors. The spatial resolution of the ALFA detectors is assumed to be of 30 $\mu$m in $x$ and $y$.

The second considered system is the AFP (ATLAS Forward Proton) detector -- horizontally moving stations planned to be installed symmetrically with respect to the ATLAS IP (IP1) at 204 m and 212 m~\cite{AFP}. Stations located closer to the IP1 will contain the tracking detectors, whereas the outer ones will be equipped with the tracking and timing devices. The reconstruction resolution of tracking detectors is foreseen to be of 10 and 30 $\mu$m in $x$ and $y$, correspondingly.

There are several LHC machine set-ups, at which the ALFA and AFP detectors could take data. In the simplest possible way they could be characterized by the value of the betatron function at the Interaction Point, $\beta^*$. In this work two such LHC machine settings will be considered: $\beta^* = 0.55$~m and $\beta^* = 90$~m. The details of these optics are described in Ref.~\cite{LHC_optics}, whereas here only the key features are presented. 

The $\beta^* = 0.55$~m is a common setting for all LHC high luminosity runs -- the beam is strongly focused at the IP and the non-zero value of the crossing angle is introduced in order to avoid collisions of proton bunches outside the IP region. The $\beta^* = 90$~m optics was developed in order to measure the properties of the elastic scattering. Due to the high value of the betatron function the beam angular divergence is very low and the beam is not as strongly focused as in case of the collision optics. In these settings the value of the crossing angle could be zero or non-zero, depending on the number of bunches. Nevertheless, as was shown in Ref.~\cite{LHC_optics}, the crossing angle has a marginal impact on the detector acceptance. Therefore, in the following this effect will not be considered.

\begin{figure}
  \centering
  \begin{subfigure}[AFP, $\beta^* = 0.55$~m]{
    \includegraphics[width=0.47\columnwidth]{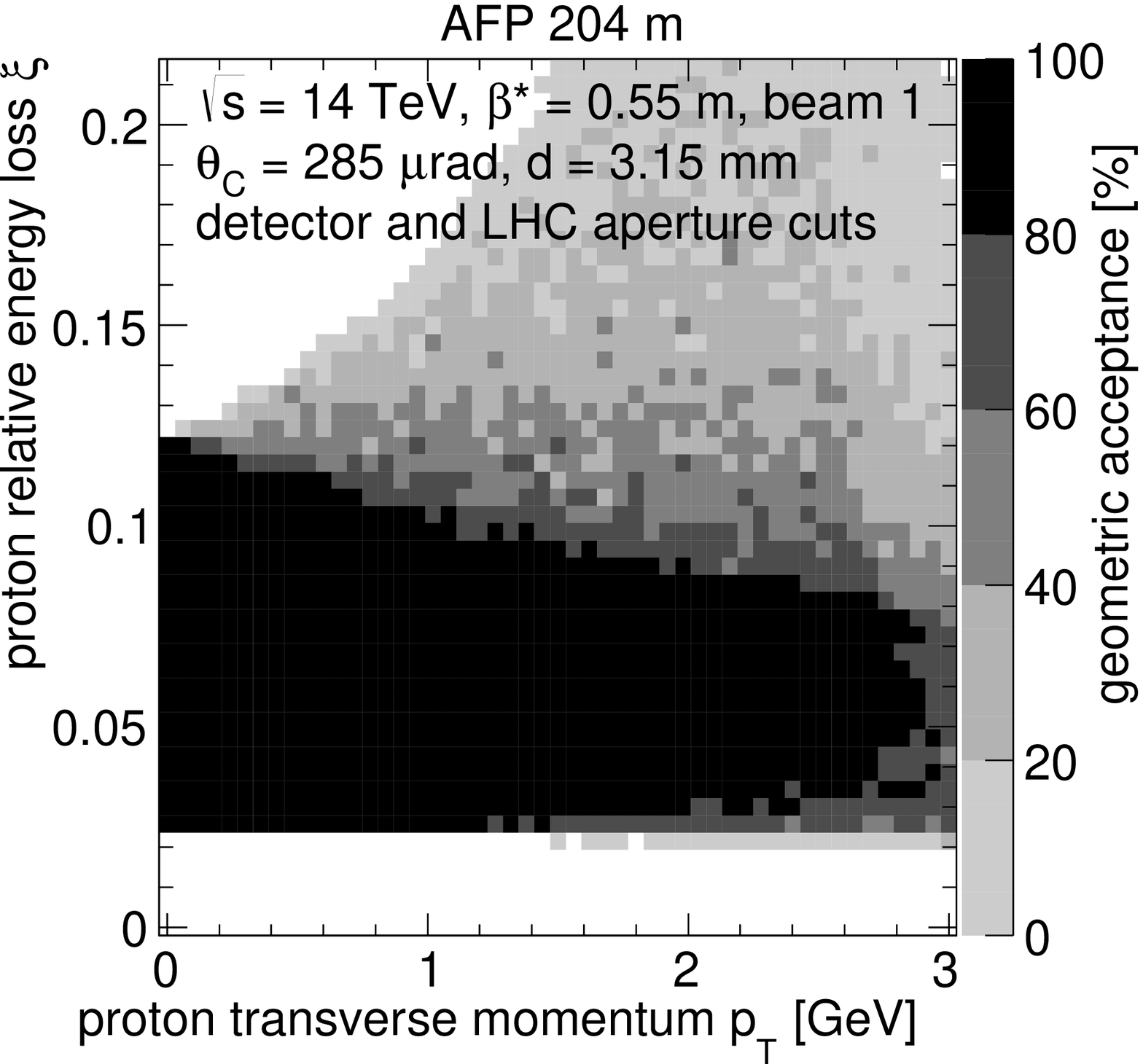}}
  \end{subfigure}
  \begin{subfigure}[AFP, $\beta^* = 90$~m]{
    \includegraphics[width=0.47\columnwidth]{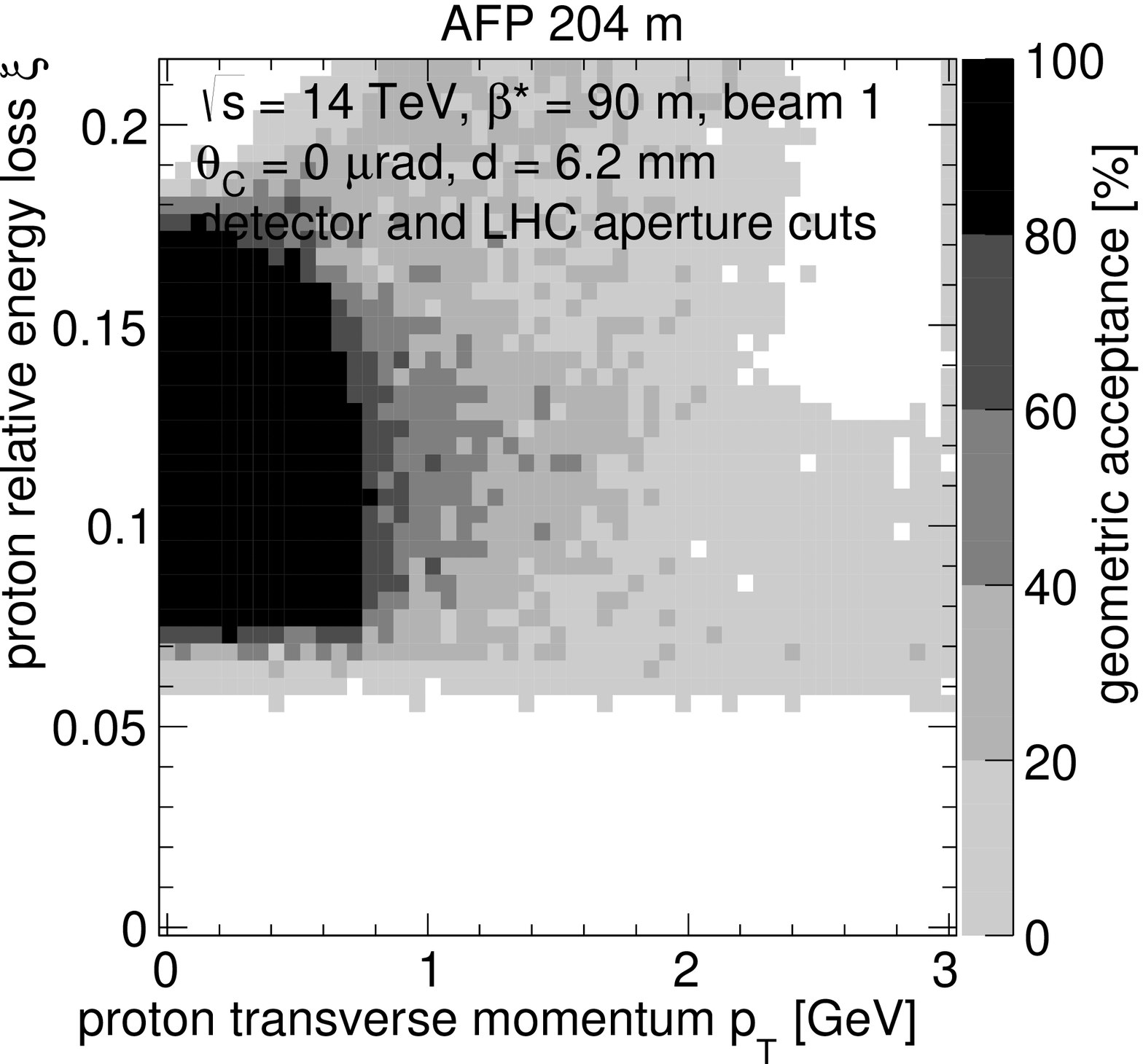}}
  \end{subfigure}
  \begin{subfigure}[ALFA, $\beta^* = 0.55$~m]{
    \includegraphics[width=0.47\columnwidth]{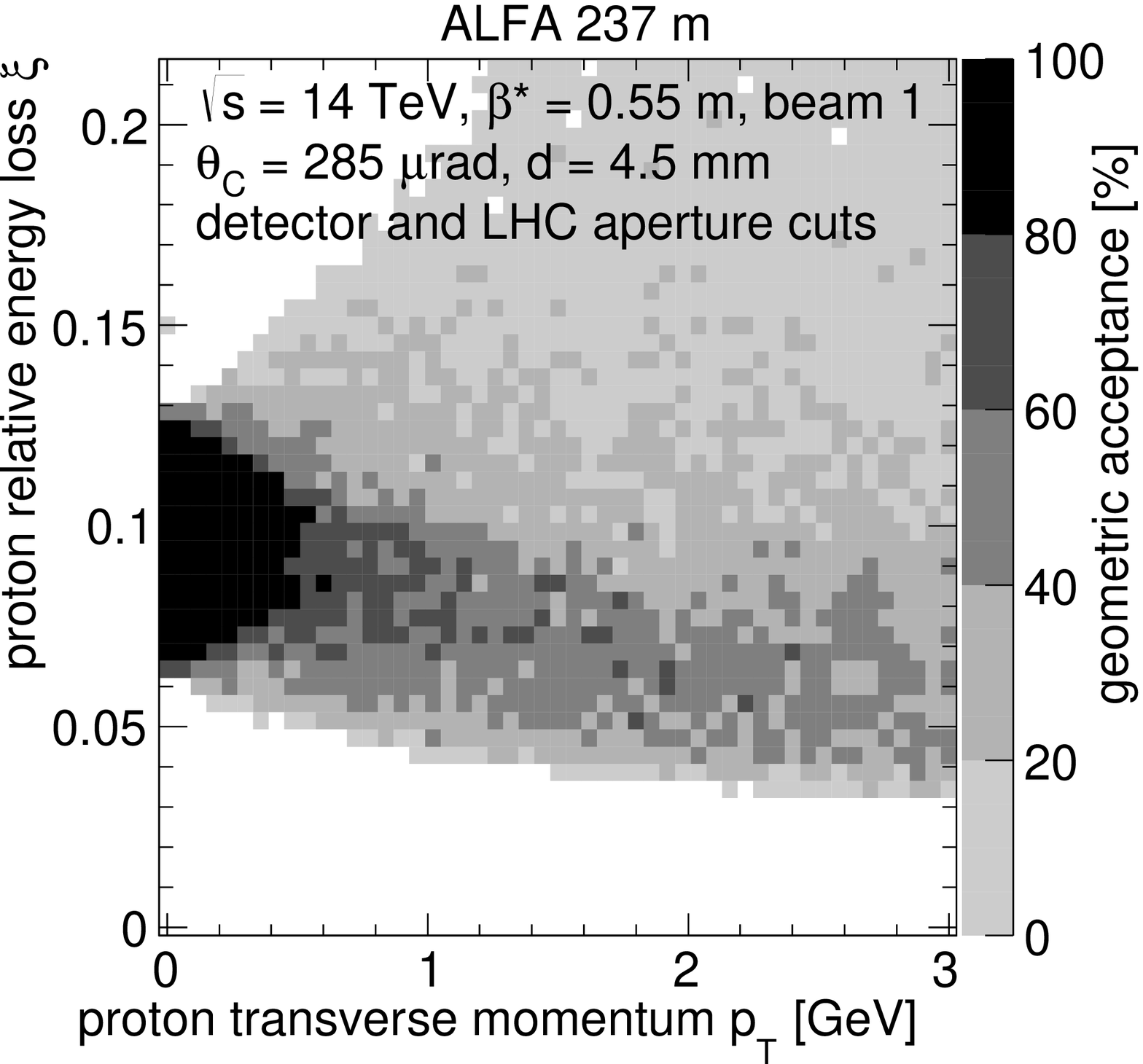}}
  \end{subfigure}
  \begin{subfigure}[ALFA, $\beta^* = 90$~m]{
    \includegraphics[width=0.47\columnwidth]{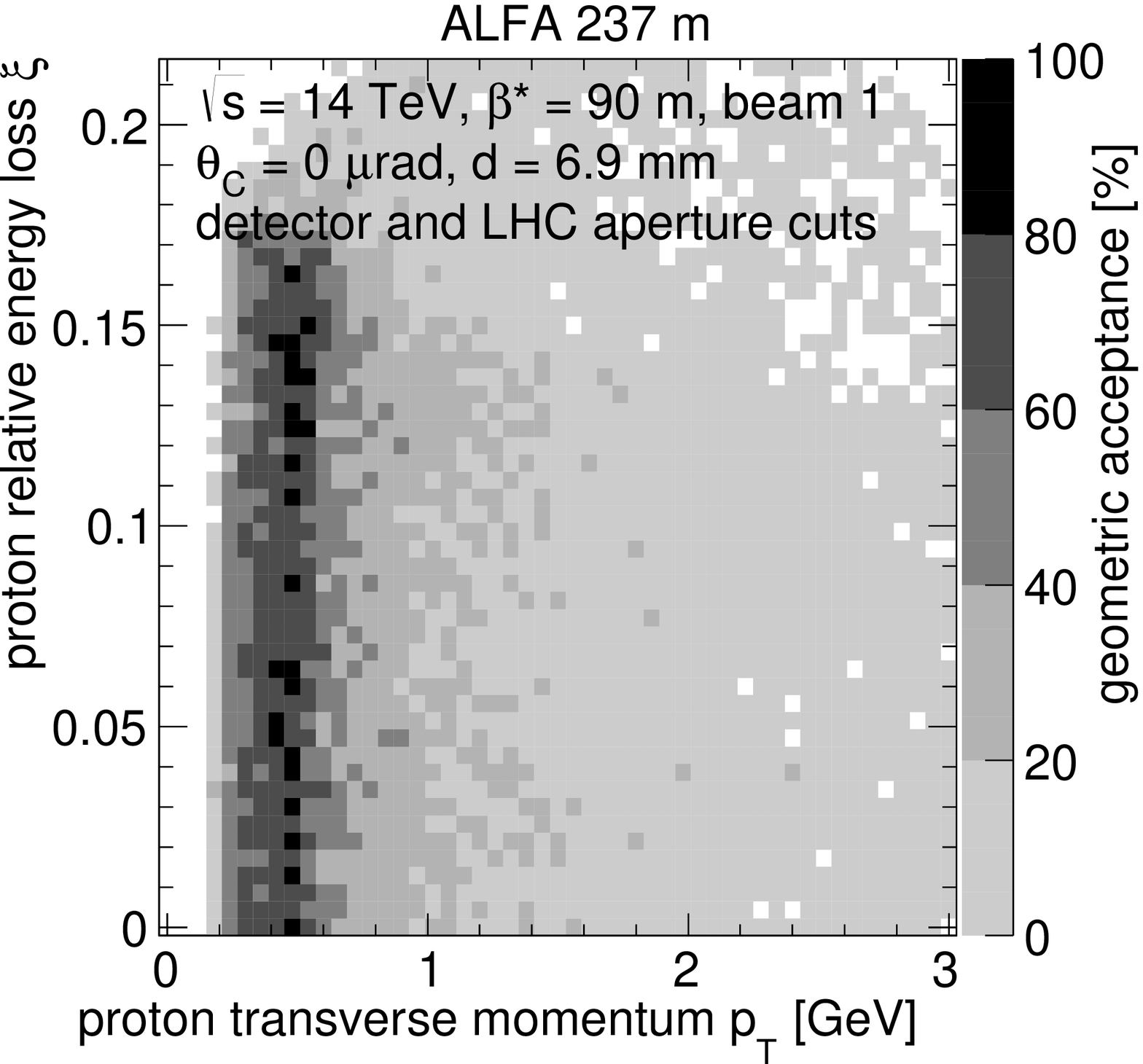}}
  \end{subfigure}
\caption{Geometric acceptance. The distance from the beam centre was set to 15 $\sigma$ for the collision optics, to 10 $\sigma$ for the \textit{high-}$\beta^*$ ones and 0.3 mm of dead material was assumed.}
\label{fig_acceptance}
\end{figure}

Not all the scattered protons can be registered in the forward detectors. A proton can be too close to the beam to be detected or it can hit the LHC element (collimator, beam pipe, magnet) upstream the AFP or ALFA station. The geometric acceptance, defined as the ratio of the number of protons of a given relative energy loss ($\xi = 1 - \frac{E_{proton}}{E_{beam}}$) and transverse momentum ($p_T$) that reached the detector station to the total number of the scattered protons having $\xi$ and $p_T$, is shown in Fig.~\ref{fig_acceptance}. In the calculations, the beam properties at the IP, the beam chamber geometry, the LHC lattice magnetic properties and the distance between the beam centre and the detector edge were taken into account. The detector distance from the beam centre was set to 15 $\sigma$ for the collision optics, to 10 $\sigma$ for the $\beta^* = 90$~m, where $\sigma$ is the beam size at the location of the detector station. Following Ref.~\cite{LHC_optics} this translates to:
\begin{itemize}
  \item 2.85 mm for AFP and $\beta^* = 0.55$~m,
  \item 5.9 mm for AFP and $\beta^* = 90$~m,
  \item 4.2 mm for ALFA and $\beta^* = 0.55$~m,
  \item 6.6 mm for ALFA and $\beta^* = 90$~m.
\end{itemize}
\noindent In order to account for the dead material of the Roman Pot window a 0.3 mm distance was added in all cases.

The geometric acceptances of ALFA and AFP detectors at various optics are complementary. For the AFP run with the $\beta^* = 0.55$~m optics the region of high acceptance (black area, $>80\%$) is limited to $p_T < 3$ GeV and $0.02 < \xi < 0.12$. These limits change to $p_T < 1$ GeV and $0.07 < \xi < 0.17$ for $\beta^* = 90$~m optics. For the ALFA detectors and $\beta^* = 0.55$~m optics the region of high acceptance is limited by $p_T < 0.5$ GeV and $0.06 < \xi < 0.12$, which is significantly smaller than in case of the AFP detectors. The picture changes drastically when $\beta^* = 90$ optics is considered, as these settings are optimised for the elastic scattering measurement in which the access to low $p_T$ values for $\xi = 0$ is crucial (\textit{c.f.} Ref.~\cite{ALFA_elastic_7TeV}).

The acceptance could be also expressed in terms of the so-called missing mass, $M_X = \sqrt{s \cdot \xi_1 \cdot \xi_2}$, where $\xi_1$ and $\xi_2$ denote the relative energy loss of the intact protons. In order to generate the mass spectrum a toy model was used:
$$\frac{\dd^4 \sigma}{\dd \xi_1\, \dd \xi_2\, \dd t_1\, \dd t_2} = \frac{\exp [b \cdot (t_1 + t_2)]}{\xi_1 \cdot \xi_2},$$
where $t_1$ and $t_2$ are the four-momentum transfer squared of the protons and the slope was set to $b = 4$ GeV$^{-2}$. Such model is expected to work for diffractive events, whereas for the exclusive ones there could be some differences (especially in $\xi$ distribution). However, this will have only small impact on the normalisation while the acceptance range will not be influenced. This is due to the fact that the distribution of the boost of the central system is qualitatively similar for all sensible models -- symmetric and flat around 0.

\begin{figure}
  \centering
  \includegraphics[width=1.0\columnwidth]{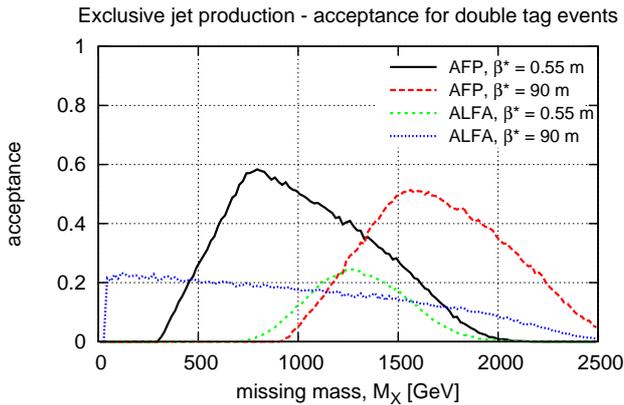}
\caption{The acceptance for events with both protons in the forward detectors as a function of the missing mass.}
\label{fig_acceptance_mass_double_tag}
\end{figure}

The acceptance for double tagged events as a function of the missing mass is shown in Figure~\ref{fig_acceptance_mass_double_tag} for both detectors and both LHC optics. The various ranges of masses are available for these settings, namely:
\begin{itemize}
  \item $300 < M_X < 1800$ GeV for AFP and $\beta^* = 0.55$~m,
  \item $1000 < M_X < 2500$ GeV for AFP and $\beta^* = 90$~m,
  \item $40 < M_X < 2500$ GeV for ALFA and $\beta^* = 0.55$~m,
  \item $800 < M_X < 1700$ GeV for ALFA and $\beta^* = 90$~m.
\end{itemize}

\begin{figure}
  \centering
  \includegraphics[width=1.0\columnwidth]{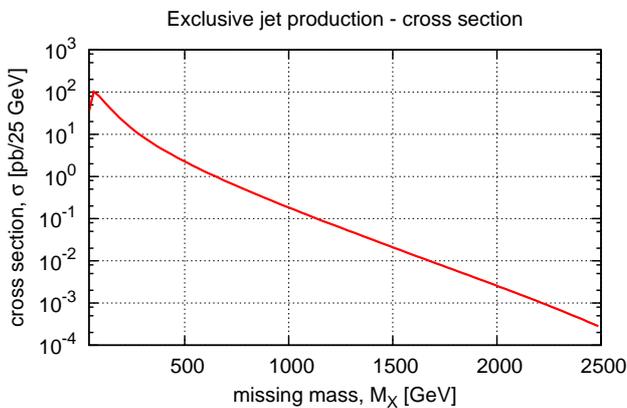}
\caption{Exclusive jet production cross section as a function of the missing mass.}
\label{fig_mass_cross_section}
\end{figure}

In order to judge whether the measurement is possible for a given experimental conditions the expected cross-section needs to be considered. Its values, as a function of missing mass, are shown in Figure~\ref{fig_mass_cross_section}.

The cross-section drops rapidly with increasing the missing mass value. Such relatively small cross-section for all cases except ALFA and $\beta^* = 90$~m optics, implies the need of large values of the integrated luminosity -- at least of the order of inverse femtobarns. In consequence, the measurement has to be performed in harsh experimental conditions where several proton-proton interactions during one bunch crossing are possible. As was shown in Ref.~\cite{ATLAS_exclusive} such a measurement is possible but very challenging. Moreover, as the signal-to-background ratio is not expected to be large, the systematic effects like the background modelling need to be considered.

\begin{figure}
  \centering
  \includegraphics[width=1.0\columnwidth]{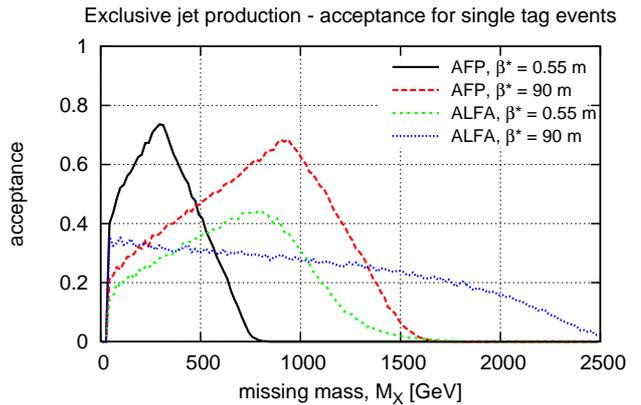}
\caption{The acceptance for events with exactly one proton in the forward detector as a function of the missing mass.}
\label{fig_acceptance_mass_single_tag}
\end{figure}

In order to remove these drawbacks, in this paper we discuss cases in which exactly one proton is within the detector acceptance, whereas the other one is too close to the beam to be detected. Such events will be hereafter named the single tagged events. The acceptance for such events as a function of the missing mass is shown in Figure~\ref{fig_acceptance_mass_single_tag} for both detectors and both LHC optics. Comparison between Figures \ref{fig_mass_cross_section} and \ref{fig_acceptance_mass_single_tag} leads to the conclusion that, except for the case of ALFA and $\beta^* = 90$~m optics, the acceptance for single tagged events is shifted towards the lower masses, hence higher cross-sections.

\section{Backgrounds}

\begin{figure}
  \centering
  \begin{subfigure}[]{
    \includegraphics[width=0.25\columnwidth]{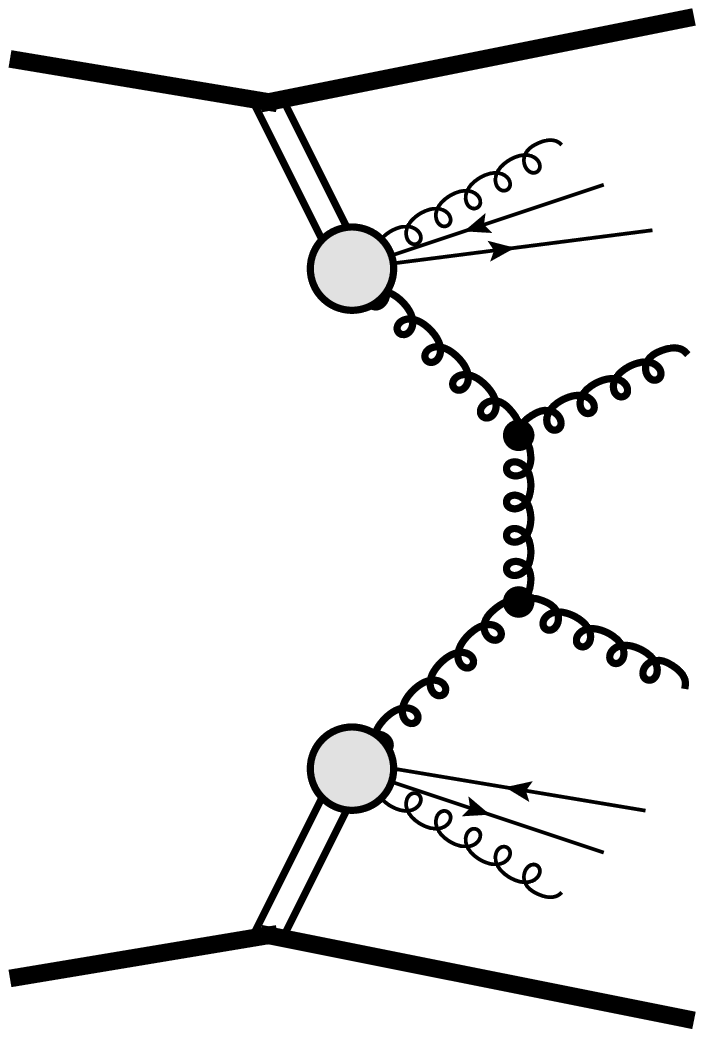}}
  \end{subfigure}
  \begin{subfigure}[]{
    \includegraphics[width=0.25\columnwidth]{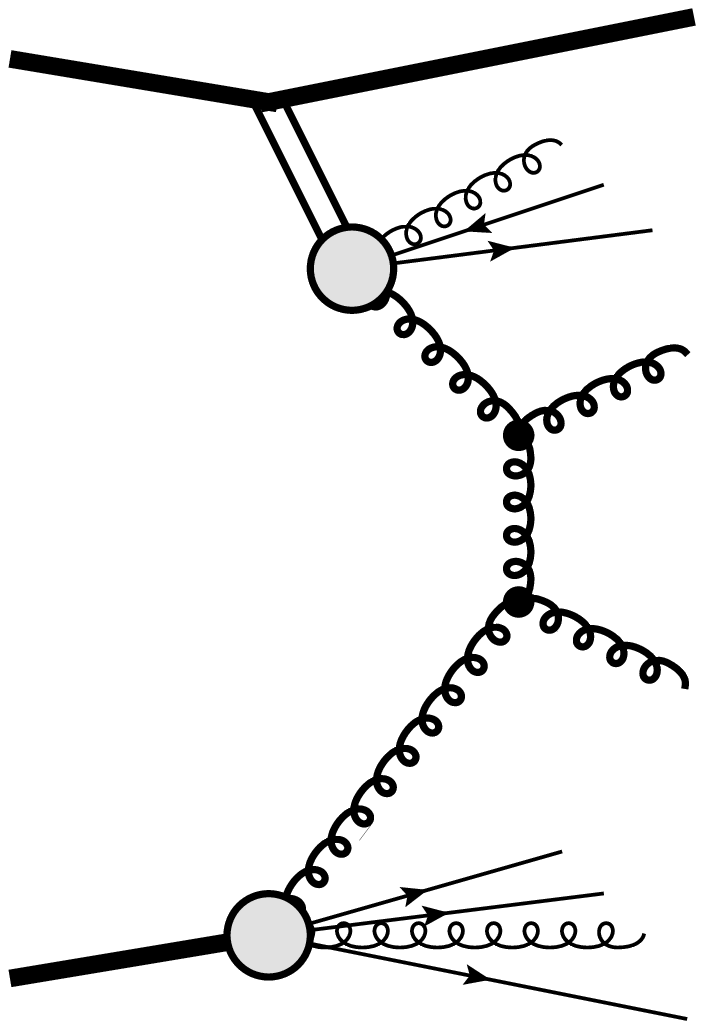}}
  \end{subfigure}
  \begin{subfigure}[]{
    \includegraphics[width=0.25\columnwidth]{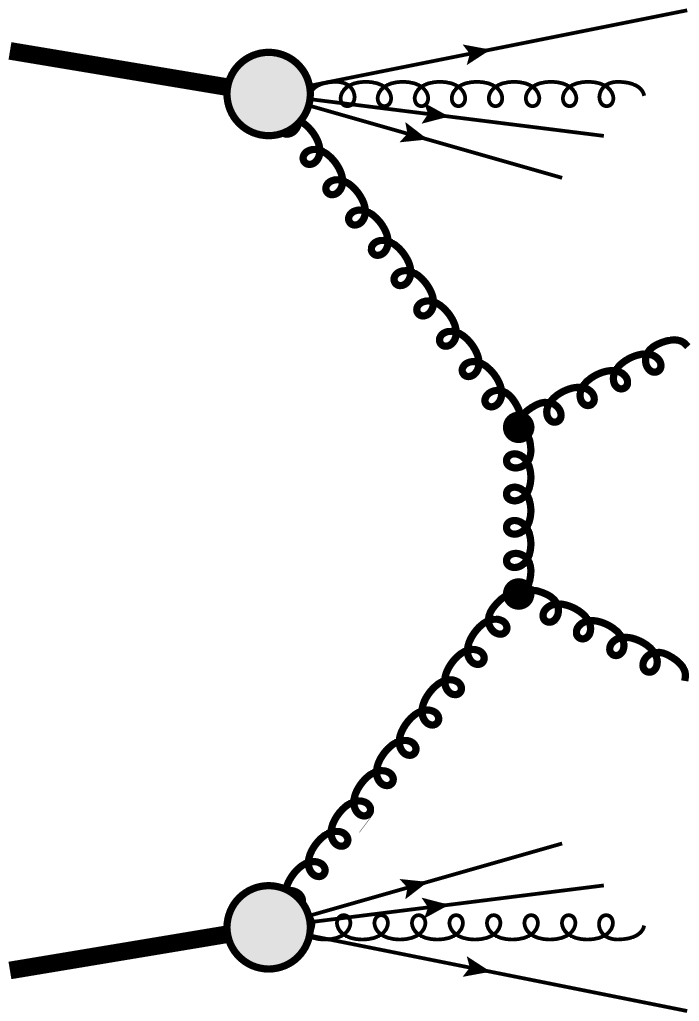}}
  \end{subfigure}
  \caption{Diagrams of background events: double Pomeron exchange (a), single diffractive (b) and non-diffractive (c) jet production. The double line marks the Pomeron exchange.}
\label{fig_bg_diag}
\end{figure}

In order to mimic a single tagged exclusive event, there has to be, apart from two jets, a final state proton visible in the forward detector. In case of the DPE jet production (see Fig.~\ref{fig_bg_diag} (a)) one of the protons has to be within and one outside the acceptance. For the single diffractive jets (Fig.~\ref{fig_bg_diag} (b)) the diffractive proton must be visible. In case of the non-diffractive jet production (Fig.~\ref{fig_bg_diag} (c)) there is no forward proton present in the system. Nevertheless, due to the non-zero pile-up value it could happen that there is simultaneous production of soft diffractive event and a diffractively scattered proton will reach the forward detector.

\section{Signal Selection}
Due to the Sudakov~\cite{Sudakov} form factor, the cross-section for the exclusive jet production is much smaller than that for the non-diffractive or diffractive jets. In consequence, in order to have a pure sample there is a need to impose several selection requirements.

\subsection*{Proton Tag}
The presence of the forward proton is a natural requirement in the presented analysis. The probabilities of observing single tagged events for both considered detectors and both optics settings are listed in Table~\ref{tab_tag_prob}. As earlier, the distance between the detector and the beam centre was set to 15~$\sigma$ and 10~$\sigma$ for the $\beta^* = 0.55$~m and the $\beta^* = 90$~m optics, correspondingly. For all detectors a presence of 0.3 mm thick layer of dead material was assumed. The values are due to the geometric acceptance (\textit{cf.} Fig~\ref{fig_acceptance}).

\begin{table}[htbp]
\caption{Probability of observing a single tagged minimum-bias event in ALFA and AFP detectors for $\beta^* = 0.55$~m and $\beta^* = 90$~m optics.}
\label{tab_tag_prob}	
  \begin{center}
    \begin{tabular}{l | c}
\toprule
\multicolumn{1}{c |}{\multirow{2}{*}{settings}} & single tag \\
\mbox{\ } & probability [\%]\\
\midrule
AFP, $\beta^* = 0.55$~m & 1.7 \\
AFP, $\beta^* = 90$~m & 1.2 \\
ALFA, $\beta^* = 0.55$~m & 0.73 \\
ALFA, $\beta^* = 90$~m & 12 \\
\bottomrule 
    \end{tabular}
  \end{center}
\end{table}

It is worth stressing that the differences between various MC generators are known to be significant and even a factor of 2 in the predicted cross-sections can be expected~\cite{phd_thesis}. However, these cross-sections should be measured at LHC prior to the exclusive jet measurement.

\subsection*{One Vertex}
In order to reduce the background originating from non-diffractive jets events, exactly one vertex reconstructed in the central detector was required. There are two main sources of interaction vertex reconstruction inefficiency:
\begin{itemize}
  \item soft event is produced too close to the hard one; due to the finite detector resolution and reconstruction algorithms the vertices are merged,
  \item there is not enough reconstructed tracks pointing to the soft vertex.
\end{itemize}
Following Ref.~\cite{ATLAS_exclusive}, the minimal distance below which vertices are merged was set to 1.5 mm.

The vertex was assumed to be reconstructed if there are at least four charged particles within $|\eta| < 2.5$ (ATLAS tracker). In order to account for the tracking efficiency each particle had a certain probability of registration/reconstruction. The thresholds were set to:
\begin{itemize}
  \item 50\% for the particles with $100 < p_T < 500$ MeV and
  \item 90\% for the ones with $p_T > 500$ MeV.
\end{itemize}
These values reflect tracking properties of the ATLAS inner detector~\cite{ATLAS_track_reco} but are also similar for the CMS experiment~\cite{CMS_track_reco}. The probabilities of observing a minimum-bias event with proton within the detector acceptance but without the reconstructed vertex are listed in Table~\ref{tab_tag_vtx_prob}.

\begin{table}[htbp]
\caption{Probability of observing a single-tagged minimum-bias event without the reconstructed vertex.}
\label{tab_tag_vtx_prob}	
  \begin{center}
    \begin{tabular}{l | c}
\toprule
\multicolumn{1}{c |}{\multirow{2}{*}{settings}} & single tag \\
\mbox{\ } & probability [\%]\\
\midrule
AFP, $\beta^* = 0.55$~m & 0.52\\
AFP, $\beta^* = 90$~m  & 0.51\\
ALFA, $\beta^* = 0.55$~m  & 0.26\\
ALFA, $\beta^* = 90$~m  & 10\\
\bottomrule 
    \end{tabular}
  \end{center}
\end{table}

The single vertex requirement will also have an impact on the exclusive, DPE and SD jet production in cases when they are produced with pile-up. In such case the probability of observing a minimum-bias vertex (w/o assumption of such an event being tagged in forward detector) must be considered. This probability is the same for all the considered settings and is of about 0.8.

\subsection*{Relative Energy Loss Difference}
In the exclusive event the relative energy loss of measured proton ($\xi_{det}$) is correlated to the jet system properties, which can be expressed as:
$$\xi_{jet} = \exp (\pm y_{jj}) \frac{M_{jj}}{\sqrt{s}},$$
where $y_{jj}$ and $M_{jj}$ are the rapidity and the mass of the jet system, respectively. In practice, it is convenient to use the relative energy loss difference:
$$\xi_{rel} = (\xi_{det} - \xi_{jet}) / (\xi_{det} + \xi_{jet}).$$

The correlation gets weaker when the detector effects are considered. The uncertainty on the $\xi_{det}$ depends on properties of a given optics and on the detector reconstruction resolution and is close to~\cite{unfolding,phd_thesis}:
\begin{itemize}
  \item 10 GeV for AFP and $\beta^* = 0.55$~m,
  \item 35 GeV for AFP and $\beta^* = 90$~m,
  \item 30 GeV for ALFA and $\beta^* = 0.55$~m,
  \item 40 GeV for ALFA and $\beta^* = 90$~m.
\end{itemize}

For jets, the uncertainty of the pseudorapidity and azimuthal angle reconstruction was assumed to be Gaussian with the width of 0.05. This value equals the doubled size of the ATLAS calorimeter cell in the central region~\cite{ATLAS}. In the calculations the effects of the Jet Energy Resolution (JER) of 20\% and the Jet Energy Scale (JES) of $\pm 10\%$ were considered. These values are conservative and based on the ATLAS detector performance~\cite{ATLAS_JER_JES}.

\begin{figure}
  \centering
   \includegraphics[width=1.0\columnwidth]{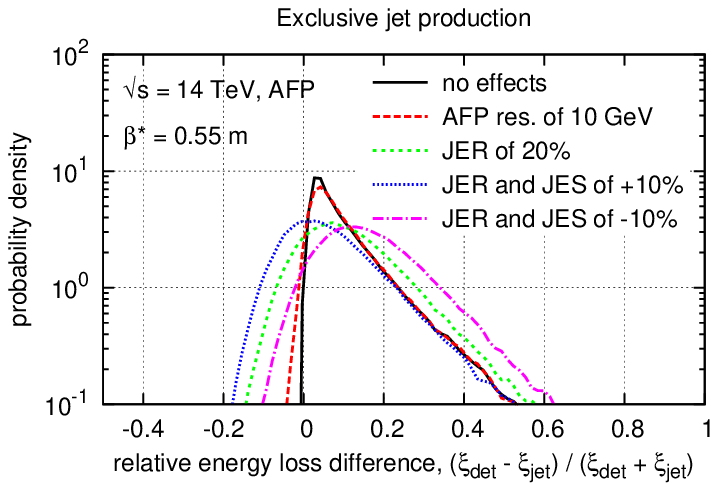}
    \includegraphics[width=1.0\columnwidth]{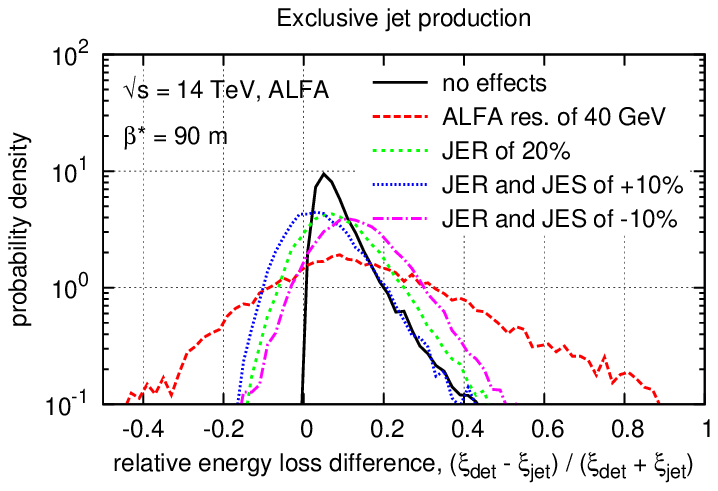}
\caption{The distribution of relative energy loss difference for AFP, $\beta^* = 0.55$~m (top) and ALFA, $\beta^* = 90$~m (bottom). The black line represents the situation when no detector effects are considered, the color ones reflect the effects of AFP $\xi$ reconstruction resolution of 10 GeV (red), JER of 20\% (green), JER of 20\% and JES of 20\% (blue), JER of 20\% and JES of -20\% (magenta).}
\label{fig_exc_rap}
\end{figure}

The results obtained for the case of the AFP detectors and $\beta^* = 0.55$~m optics are presented in Fig.~\ref{fig_exc_rap} (top). The black lines represent the situation when no detector effects are taken into account. The tail in case of \textit{no effects} is due to the presence of additional activity in the system (\textit{e.g.} third jet). The effects of inaccurate $\xi$ reconstruction from the detector measurements (red lines) are much smaller than those due to the Jet Energy Resolution (green lines). The effects of Jet Energy Scale shift the distributions towards the higher (JES of -10\%, magenta lines) or lower (JES of 10\%, blue lines) values. The effects due to JER and JES are much more important that the ones due to the AFP detector resolution.

For AFP, $\beta^* = 90$~m and ALFA, $\beta^* = 0.55$~m the shapes of the relative energy loss difference distributions are similar to the AFP and $\beta^* = 0.55$~m case. For ALFA and $\beta^* = 90$~m the smearing is mainly due to the detector resolution (see Fig.~\ref{fig_exc_rap} (bottom)).

\begin{figure}
  \centering
   \includegraphics[width=1.0\columnwidth]{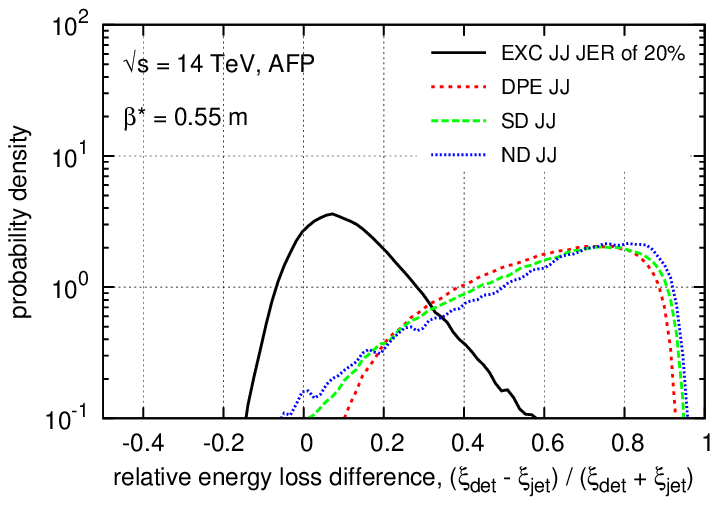}
   \includegraphics[width=1.0\columnwidth]{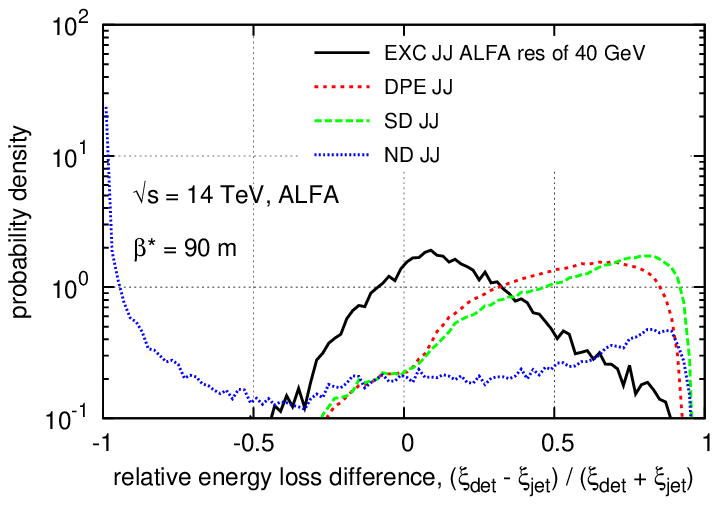}
\caption{The distribution of relative energy loss difference for AFP, $\beta^* = 0.55$~m (top) and ALFA, $\beta^* = 90$~m (bottom). The black solid line is for the signal (taking into account smearing due to JER), whereas the other ones represent backgrounds: double Pomeron exchange (red), single diffractive (gren) and non-diffractive (blue) jets.}
\label{fig_bkg_rap}
\end{figure}

The distribution of the relative energy loss difference for signal and background events -- AFP, $\beta^* = 0.55$~m -- is shown in Fig.~\ref{fig_bkg_rap} (top). The black solid line marks the signal distribution (taking into account the smearing due to JER), whereas the other ones represent the backgrounds: double Pomeron exchange (red), single diffractive (gren) and non-diffractive (blue) jets. The signal becomes smaller than the background for $\xi_{rel} > 0.3$ and this cut-off value is used in further analysis. For AFP, $\beta^* = 90$~m and ALFA, $\beta^* = 0.55$~m backgrounds are slightly shifted towards higher values of $\xi_{rel}$. In consequence, events with $\xi_{rel} < 0.4$ are accepted.

The result of calculations for ALFA and $\beta^* = 90$~m is shown in Fig.~\ref{fig_bkg_rap} (bottom). In this plot the black solid line represents the shape of the signal distribution. 
The uncertainty coming from the resolution of ALFA detector was taken into account.
Non-diffractive events present in the region of $\xi_{rel} < 0$ are due to acceptance for low-$\xi$ minimum-bias protons (\textit{cf.} Fig.~\ref{fig_acceptance} (d)). Considering the shapes of the signal and background distributions, events with $|\xi_{rel}| < 0.3$ were accepted.

\subsection*{Number of Tracks and Deposited Energy}
The lack of both proton/Pomeron remnants and underlying the event activity provides another handle on the improvement of the signal purity. 

\begin{figure}
  \centering
  \begin{subfigure}{
    \includegraphics[width=1.0\columnwidth]{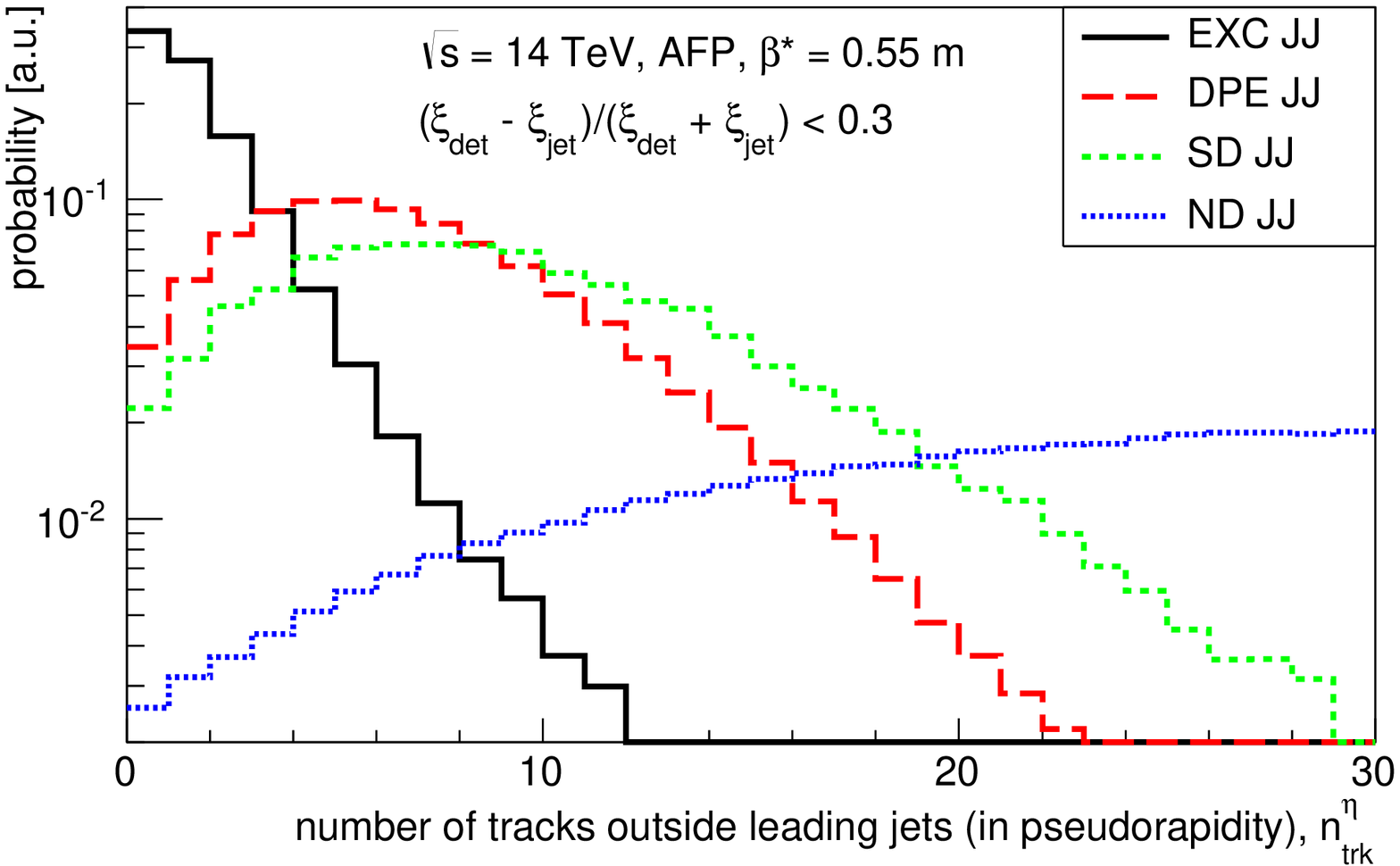}}
  \end{subfigure}
  \begin{subfigure}{
    \includegraphics[width=1.0\columnwidth]{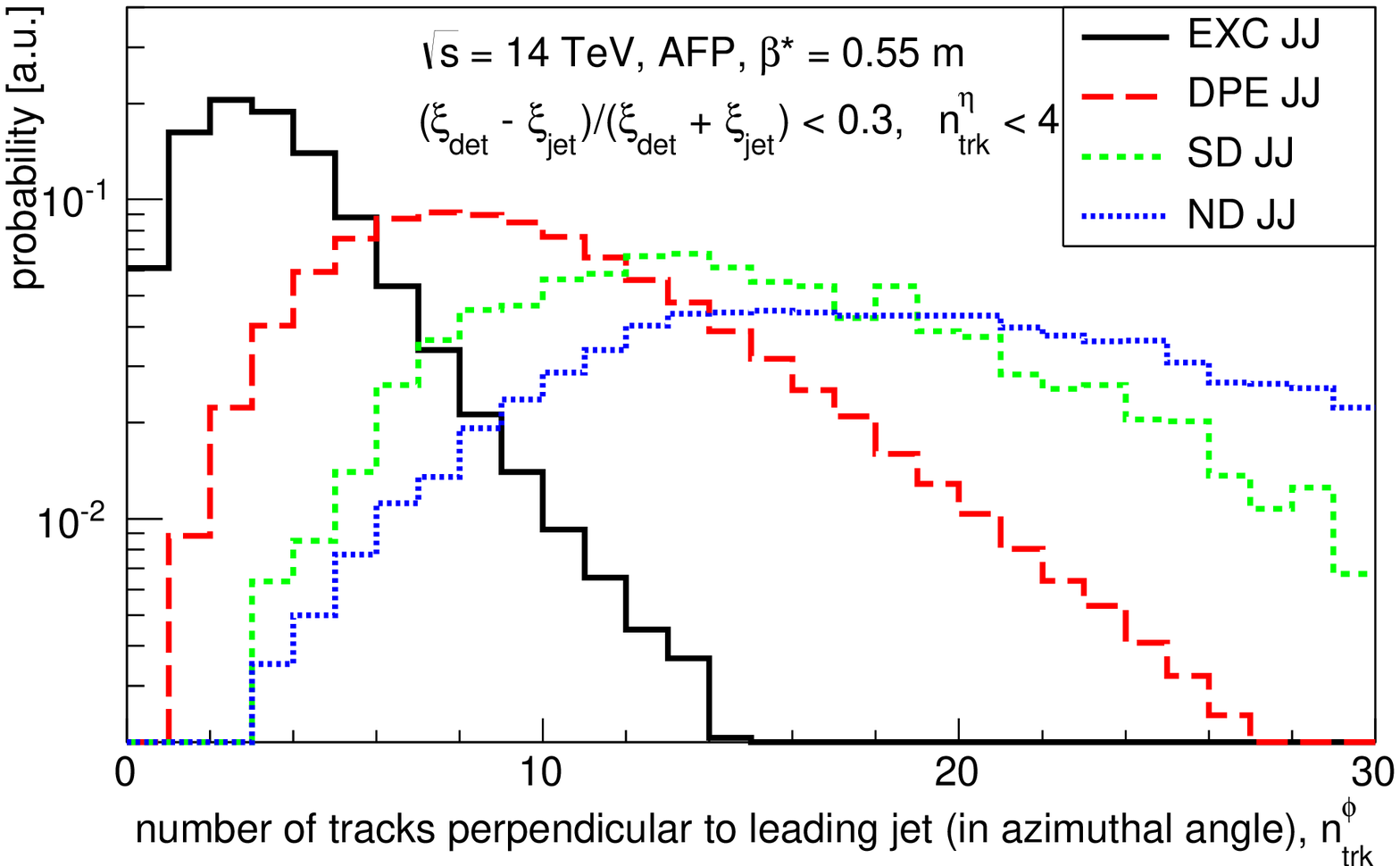}}
  \end{subfigure}
\caption{The number of tracks outside the jet system in outside the jet system in pseudorapidity ($n^{\eta}_{trk}$, top) and the number of tracks perpendicular to the leading jet in azimuthal angle ($n^{\phi}_{trk}$, bottom) for the signal and background events. The integral of the distribution is normalised to 1 (overflow bins are considered).
}
\label{fig_cut_ntrk}
\end{figure}

The distributions of the number of tracks outside the jet system in pseudorapidity, $n^{\eta}_{trk}$, and the number of tracks perpendicular to the leading jet in azimuthal angle, $n^{\phi}_{trk}$, are shown in Fig.~\ref{fig_cut_ntrk} (top) and (bottom), correspondingly. Since these distributions are similar for all considered settings, only results for the AFP and $\beta^* = 0.55$~m are shown as an example.

A track is considered to be outside the jet system if $\eta_{trk} > \eta_{jet}^{+} + 0.6$, where $\eta_{jet}^{+}$ is a direction of the jet (leading or sub-leading) with the highest pseudorapidity. This equation is valid only for tracks with $\eta_{trk} > 0$, but the requirement for the other case is trivial. For all considered data-taking scenarios events were accepted if $n^{\eta}_{trk} < 4$.

A track was considered as perpendicular to the leading jet if $\frac{\pi}{3} < \Delta\phi < \frac{2\pi}{3}$ or $\frac{4\pi}{3} < \Delta\phi < \frac{5\pi}{3}$, where $\Delta\phi$ is the difference between the azimuthal angle of the track and the leading jet. The tail of the track multiplicity for signal events is due to the final state parton showering and hadronisation. For all considered data-taking scenarios events were accepted if the $n^{\eta}_{trk}$ requirement is fulfilled and $n^{\phi}_{trk} < 6$.

\begin{figure}
  \centering
    \includegraphics[width=1.0\columnwidth]{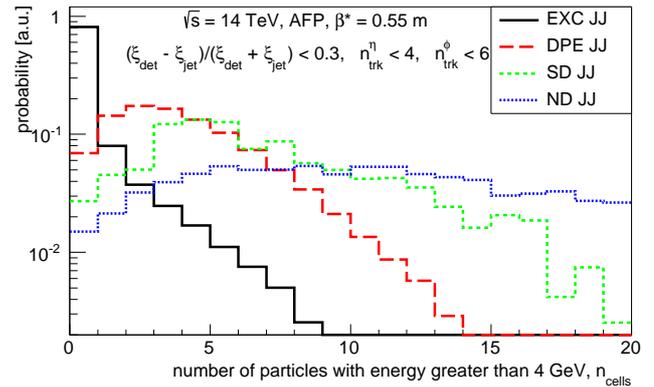}
\caption{The number of particles with the energy greater than 4 GeV produced in pseudorapidity range of $2.5 < |\eta| < 4.9$. The integral of the distribution is normalised to 1 (overflow bins are considered).}
\label{fig_cut_energy}
\end{figure}

Apart from the veto on the activity in the tracker, information coming from forward calorimeters can be also used. The distribution of the multiplicity of particles with energy greater than 4~GeV produced in pseudorapidity range of $2.5 < |\eta| < 4.9$, $n_{cells}$, is shown in Fig.~\ref{fig_cut_energy}. The 4 GeV threshold is well above the calorimeter noise~\cite{ATLAS_noise}. Event was accepted if $n_{cells} < 2$.

\subsection*{Visible Cross-section}
\begin{figure}
  \centering
    \includegraphics[width=1.0\columnwidth]{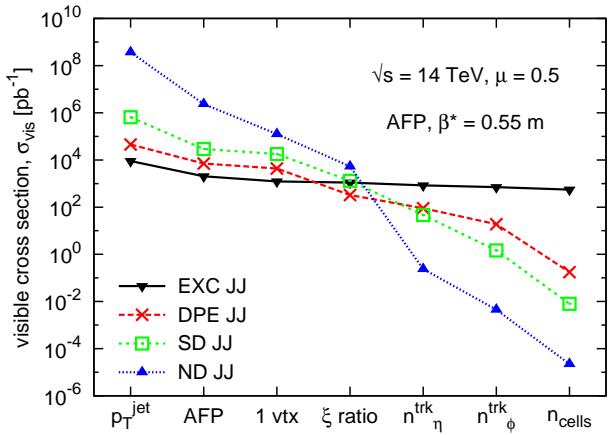}
\caption{The number of events accepted after a particular cut for signal and background for an average number of interactions of $\mu = 0.5$ as a function of the applied consecutive cuts for AFP and $\beta^* = 0.55$~m.}
\label{fig_cut_summary}
\end{figure}

The change of the visible cross-section value for the signal and background processes after each selection requirement for AFP, $\beta^* = 0.55$~m and the average number of interactions of $\mu = 0.5$ is visualised in Fig.~\ref{fig_cut_summary}. The black solid line marks the signal, whereas the other ones represent the backgrounds: DPE (red), SD (green) and ND (blue) jets. After all the selection requirements the signal to background ratio increases from $10^{-5}$ to the final value of $10^4$.

\begin{figure}
  \centering
    \includegraphics[width=1.0\columnwidth]{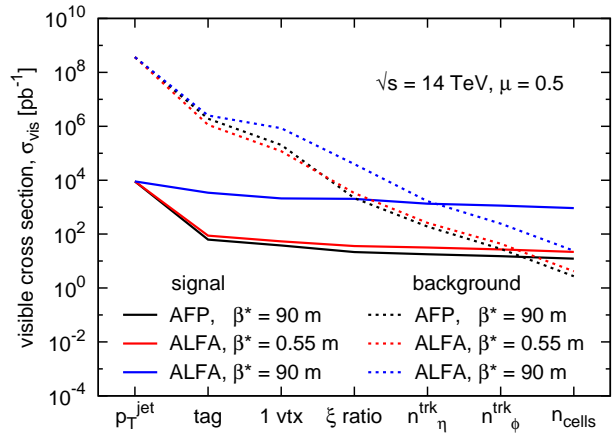}
\caption{The number of events accepted after a particular cut for signal and background for an average number of interactions of $\mu = 0.5$ as a function of the applied consecutive cuts. Black lines are for AFP and $\beta^* = 90$~m, red for ALFA and $\beta^* = 0.55$~m and blue for ALFA and $\beta^* = 90$~m.}
\label{fig_cut_summary_all}
\end{figure}

The results for other considered experimental set-ups are shown in Fig.~\ref{fig_cut_summary_all}. The solid line represents the signal whereas the dotted ones mark the sum of the DPE, SD and ND backgrounds. The black lines are for AFP and $\beta^* = 0.55$~m, the red -- for AFP and $\beta^* = 0.55$~m and the blue ones for ALFA and $\beta^* = 90$~m. For ALFA and $\beta^* = 90$~m the signal-to-background ratio is $\sim 40$. For the other two data-taking scenarios this ratio is about 5. This is mainly due to the smaller ALFA detector acceptance in the region of low missing mass (\textit{cf.} Fig.~\ref{fig_acceptance_mass_single_tag}). It is worth stressing that since the background is mainly due to the single diffractive and double Pomeron exchange jet productions, the signal-to-background ratio will not change significantly for $\mu \lesssim 5$.

On the basis of the results shown in Figs~\ref{fig_cut_summary} and \ref{fig_cut_summary_all} the expected number of events can be estimated. For example, assuming the integrated luminosity equal to 10~pb$^{-1}$, the expected number of events is of about:
\begin{itemize}
  \item 4000 for AFP and $\beta^* = 0.55$~m,
  \item 120 for AFP and $\beta^* = 90$~m,
  \item 210 for ALFA and $\beta^* = 0.55$~m,
  \item 9000 for ALFA and $\beta^* = 90$~m.
\end{itemize}
The value of 10~pb$^{-1}$ was chosen as possible to be obtained during low-luminosity runs at the LHC.

\begin{figure}
  \centering
    \includegraphics[width=1.0\columnwidth]{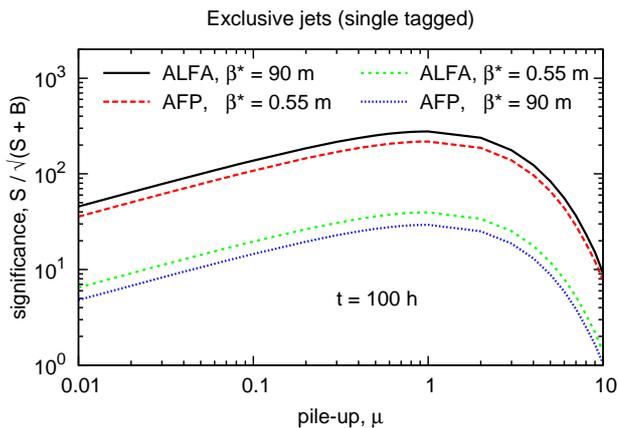}
\caption{Statistical significance for 100 hours of data-taking as a function of pile-up.}
\label{fig_significance}
\end{figure}

The quality of the measurement can be expressed in terms of the statistical significance defined as $\frac{S}{\sqrt{S + B}}$, where $S$ and $B$ are numbers of the collected signal and background events, correspondingly. The statistical significance for all the considered scenarios and the data-taking time of 100 hours as a function of pile-up is shown in Fig.~\ref{fig_significance}. The black line represents the distribution for the ALFA detectors and $\beta^* = 90$~m optics, the red one is for AFP and $\beta^* = 0.55$~m, green -- for ALFA and $\beta^* = 0.55$~m and the blue one for AFP and $\beta^* = 90$~m. For all cases the maximal significance is obtained for pile-up of about~1. Slow decrease of significance for $\mu < 1$ is due to the amount of data that could be collected during the fixed time, whereas the decrease for $\mu > 1$ follows from the single vertex requirement. For the considered selection criteria, the most significant measurements can be done with AFP, $\beta^* = 0.55$~m and ALFA, $\beta^* = 90$~m settings. For the other two scenarios the significance is about an order of magnitude smaller.

\section{Summary and Conclusions}
The measurement of central exclusive jet production using proton tagging technique was discussed. The studies were performed at the centre-of-mass energy of $\sqrt{s}$ = 14~TeV and the ATLAS detector. Four data-taking scenarios were considered: AFP and ALFA detectors as forward proton taggers and $\beta^* = 0.55$~m, $\beta^* = 90$~m optics. The measurement was proven to be feasible. The results are also applicable for the CMS-TOTEM experiments. 

The cross-section for the exclusive jet production drops rapidly with increasing the missing mass of the produced system. For the considered data-taking scenarios (except ALFA and $\beta^* = 90$~m optics) the requirement of both protons being within the acceptance of the forward detectors leads to a substantial decrease of the visible cross-section and hence implies large values of the integrated luminosity -- at least of the order of inverse femtobarns. In consequence, the measurement has to be performed in harsh experimental conditions where several proton-proton interactions can happen during one bunch crossing.

In order to reach the region of lower missing masses (and higher cross-sections) an analysis based on single tagged events was performed. After dedicated signal selection cuts have been applied, the signal-to-background ratio increases from $10^{-5}$ to between $5$ and $10^4$ depending on the considered run scenario. This means that the analyses of signal properties can be performed in much cleaner experimental environment. Moreover, significant measurements can be carried out for data collection period of about $100$~h with pile-up of about~$1$.

\end{document}